\documentclass[article,aps,nofootinbib,showpacs,amsfonts,epsf]{revtex4}

\input{epsf.tex}

\newcommand{\E}{{\cal{E}}}

\newcommand{\s}{\sigma}

\renewcommand{\d}{{\rm d}}
\renewcommand{\a}{\alpha}

\newcommand{\dfrac}[2]{\displaystyle\frac{#1}{#2}}
\newcommand{\be}{\begin{equation}}
\newcommand{\ee}{\end{equation}}
\newcommand{\bea}{\begin{eqnarray}}
\newcommand{\eea}{\end{eqnarray}}
\newcommand{\ba}{\begin{array}}
\newcommand{\ea}{\end{array}}
\def\J#1#2#3#4{{#1} {\bf #2}, #3 (#4)}
\def\PRD{Phys. Rev. D}

\def\APL{Ann. Phys. (Leipzig)}

\def\CQG{Class. Quantum Grav.}
\def\JPA{J. Phys. A: Math. Gen.}
\def\ZP{Z. Phys.}

\begin{document}
\draft

\title{A combined Majumdar--Papapetrou--Bonnor field\\as extreme limit of the double--Reissner--Nordstr\"om solution}

\author{I.~Cabrera--Munguia,$^\dag$ V.~S.~Manko,$^\dag$ E.~Ruiz$\,^\ddag$}
\address{$^\dag$Departamento de F\'\i sica, Centro de Investigaci\'on y de
Estudios Avanzados del IPN, A.P. 14-740, 07000 M\'exico D.F.,
Mexico\\$^\ddag$Instituto Universitario de F\'{i}sica
Fundamental y Matem\'aticas, Universidad de Salamanca, 37008 Salamanca, Spain}

\begin{abstract}
The general extreme limit of the double--Reissner--Nordstr\"om solution is worked out in explicit analytical form involving prolate spheroidal coordinates. We name it {\it the combined Majumdar--Papapetrou--Bonnor field} to underline the fact that it contains as particular cases the two--body specialization of the well--known Majumdar--Papapetrou solution and Bonnor's three--parameter electrostatic field. To the latter we give a precise physical interpretation as describing a pair of non--rotating extremal black holes with {\it unequal} masses and {\it unequal} opposite charges kept apart by a strut, the absolute values of charges {\it exceeding} the respective (positive) values of masses. \end{abstract}


\maketitle


\section{Introduction}

The Majumdar--Papapetrou (MP) \cite{Maj,Pap} electrostatic solution of the Einstein--Maxwell equations can describe in the axisymmetric case a system of aligned non--rotating charged masses which are in neutral equilibrium due to balance of the gravitational and electric forces. Although it was first thought that this solution ``is not of great physical interest''~\cite{Bon}, later on the well--known paper by Hartle and Hawking \cite{HHa}, in which the charged sources of the MP spacetime located on the symmetry axis were shown to be the extreme Reissner--Nordstr\"om (RN) black holes \cite{Rei,Nor}, quickly converted the MP solution into an important ingredient of the modern black--hole physics, with some distinctive thermodynamical properties \cite{HHR}. Since the MP black--hole spacetime formally belongs to the Weyl family of electrostatic solutions known from 1917 \cite{Wey}, it is quite surprising that it was only discovered three decades after Weyl's work, besides, first in the framework of non--symmetric, conformastatic spacetimes~\cite{Boo}. This could probably be explained by the fact that already in the two--body case the performing of the extreme limit in the Weyl solution, needed for obtaining the corresponding MP metric, is technically so tangled that it has never been attempted in the literature. In this respect it would certainly be of interest to examine the limit of extreme black holes in the five--parameter double--Reissner--Nordstr\"om (DRN) solution \cite{BMA,Man} which describes the non--linear superposition of two arbitrary RN fields and hence contains the two--body electrostatic Weyl spacetime as a particular case. By considering such a limit one could achieve the following two goals: ($i$) to show how the MP solution for two extreme RN black holes naturally arises from the general DRN spacetime, and ($ii$) to identify and analyze the remaining extreme cases.

In the present paper, following the main ideas of Ref.~\cite{MSM} dealing with a pair of stationary electrovac extreme sources, we will work out the general extreme limit of the DRN solution by making use of the canonical set of parameters $\{\a_n,\beta_l\}$ in terms of which earlier in the paper \cite{BMA} were written both the general metric for $N$ aligned RN sources and its $N=2$ specialization representing two black holes. By the term `extreme', with regard to a non--isolated RN black hole, we shall mean, as usual, the marginal black--hole state previous to the formation of a naked singularity. Curiously, it turns out that the physical representation of the DRN solution obtained in \cite{Man} by introducing the Komar masses and charges \cite{Kom} as arbitrary parameters, is inappropriate for achieving such a goal since it produces hardly manageable expressions while performing the limit $\sigma_1\to0$, $\sigma_2\to0$. However, formulae relating various mathematically equivalent parameter sets that were derived in the paper \cite{Man} will be used by us for clarifying the physical meaning of the resulting extreme solutions.

Our paper is organized as follows. In Sect.~II we first provide the reader with a necessary information about the canonical parametrization of the DRN solution and then elaborate the limit of two extreme black holes, writing it down in terms of two determinants of fifth order. In Sect.~III we consider expansions of these determinants, arriving as a result at the combined Majumdar--Papapetrou--Bonnor spacetime which contains as special cases the MP two--body solution \cite{Maj,Pap} and Bonnor's three--parameter electrostatic solution \cite{Bon2}. Here we also show that the Bonnor solution is just the electrostatic analog of the well--known Kerr--NUT spacetime \cite{DNe}, thus obtainable from the latter via a complex continuation of the parameters \cite{Bon3}, and we interpret it, in the line of papers \cite{Emp,LTe}, as describing two extreme black holes with unequal masses and unequal charges (the absolute values of charges being greater than the respective positive values of masses) which are separated by a strut. Discussion and conclusions are left for Sect.~IV.

\section{The extreme DRN solution in terms of determinants}

We recall that the general DRN solution arises from the {\it axis data} (short for values of the Ernst potentials \cite{Ern} on the upper part of the symmetry axis) of the form \cite{BMA}
\be \E(\rho=0,z)=1+\frac{e_1}{z-\beta_1}+\frac{e_2}{z-\beta_2}, \quad \Phi(\rho=0,z)=\frac{f_1}{z-\beta_1}+\frac{f_2}{z-\beta_2}, \label{data} \ee
where the parameters $e_l$, $\beta_l$ and $f_l$, $l=1,2$, can take on arbitrary real values or occur in complex conjugate pairs $e_1=\bar e_2$, $\beta_1=\bar\beta_2$, $f_1=\bar f_2$ (all of them simultaneously) due to the reality of the potentials $\E$ and $\Phi$ in the electrostatic case. Note that these parameters define only five physical real quantities because the axis expressions of the Ernst potentials (\ref{data}) include an arbitrary shift along the symmetry $z$--axis which can always be fixed at some real value. The total mass $M_{tot}$ and total charge $Q_{tot}$ of the system are given by the formulas
\be M_{tot}=-(e_1+e_2)/2, \quad Q_{tot}=f_1+f_2. \label{MQ_t} \ee

The application of Sibgatullin's integral method \cite{Sib,SMa} to the axis data (\ref{data}) yields the following expressions for $\E$ and $\Phi$ in the whole space \cite{BMA}:
\bea \E&=&E_+/E_-,
\qquad \Phi=F/E_-, \nonumber\\
E_\pm&=&\left|\begin{array}{cccc} 1 & 1 & \ldots & 1 \\ \pm 1 &
{\displaystyle \frac{r_1}{\a_1-\beta_1}} & \ldots & {\displaystyle
\frac{r_{4}}{\a_{4}-\beta_1}} \vspace{0.15cm}\\ \pm 1 &
{\displaystyle \frac{r_1}{\a_1-\beta_2}} & \ldots & {\displaystyle
\frac{r_{4}}{\a_{4}-\beta_2}} \vspace{0.15cm}\\
0 & {\displaystyle \frac{h_1(\a_1)}{\a_1-\beta_1}} & \ldots &
{\displaystyle \frac{h_1(\a_{4})}{\a_{4}-\beta_1}}
\vspace{0.15cm}\\ 0 & {\displaystyle
\frac{h_2(\a_1)}{\a_1-\beta_2}} & \ldots & {\displaystyle
\frac{h_2(\a_{4})}{\a_{4}-\beta_2}}\\
\end{array}\right|, \quad
F=\left|\begin{array}{cccc} 0 & f(\a_1) & \ldots & f(\a_{4})
\\ -1 & {\displaystyle \frac{r_1}{\a_1-\beta_1}} & \ldots &
{\displaystyle \frac{r_{4}}{\a_{4}-\beta_1}} \vspace{0.15cm}\\  -1
& {\displaystyle \frac{r_1}{\a_1-\beta_2}} & \ldots &
{\displaystyle \frac{r_{4}}{\a_{4}-\beta_2}} \vspace{0.15cm}\\ 0 &
{\displaystyle \frac{h_1(\a_1)}{\a_1-\beta_1}} & \ldots &
{\displaystyle \frac{h_1(\a_{4})}{\a_{4}-\beta_1}}
\vspace{0.15cm}\\ 0 & {\displaystyle
\frac{h_2(\a_1)}{\a_1-\beta_2}} & \ldots & {\displaystyle
\frac{h_2(\a_{4})}{\a_{4}-\beta_2}}\\
\end{array}\right|, \nonumber\\ r_n&=&\sqrt{\rho^2+(z-\a_n)^2}, \quad h_l(\a_n)=e_l+2f_l f(\a_n), \quad f(\a_n)=\sum\limits_{l=1}^2\frac{f_l}{\a_n-\beta_l}, \label{EF} \eea
where instead of the initial set of parameters $\{e_l,\beta_l,f_l\}$, a mathematically equivalent parameter set $\{\a_n,\beta_l\}$ is used, the new constants $\a_n$, $n=1,2,3,4$, taking arbitrary real values or occurring in complex conjugate pairs. The form of the `old' parameters $e_l$ and $f_l$ in terms of $\a_n$ and $\beta_l$ is the following \cite{BMA}:
\bea
e_1&=&[-2\s_4+(\beta_1+\beta_2)\s_3-2\beta_1\beta_2\s_2
+\beta_1^2(3\beta_2-\beta_1)\s_1 \nonumber\\
&&+2\beta_1^3(\beta_1-2\beta_2)](\beta_1-\beta_2)^{-3}
-2f_1f_2(\beta_1-\beta_2)^{-1}, \nonumber\\
e_2&=&[2\s_4-(\beta_1+\beta_2)\s_3+2\beta_1\beta_2\s_2
-\beta_2^2(3\beta_1-\beta_2)\s_1 \nonumber\\
&&-2\beta_2^3(\beta_2-2\beta_1)](\beta_1-\beta_2)^{-3}
+2f_1f_2(\beta_1-\beta_2)^{-1}, \nonumber \\
f_1^2&=&\frac{\prod_{n=1}^4(\beta_1-\a_n)}{(\beta_1-\beta_2)^2}, \quad
f_2^2=\frac{\prod_{n=1}^4(\beta_2-\a_n)}{(\beta_1-\beta_2)^2}, \nonumber\\
\s_1&=&\sum\limits_{i=1}^4\a_i, \quad
\s_2=\sum\limits_{1\le i<j\le4}\a_i\a_j, \quad
\s_3=\sum\limits_{1\le i<j<k\le4}\a_i\a_j\a_k, \quad
\s_4=\prod\limits_{i=1}^4\a_i. \label{ef_a} \eea

The constants $\a_n$ define the location of sources on the symmetry axis. When all $\a_n$ are real, these define two black holes; the complex conjugate pairs of $\a$s determine hyperextreme constituents. Therefore, it is clear that for passing from the case of two subextreme black holes to the case of extreme black--hole constituents we should work with the real--valued $\a_n$, assigning them first of all some order; the order adopted here is
\be \a_1\le\a_2<\a_3\le\a_4. \label{order} \ee
Then the parts $\rho=0$, $\a_2\le z\le\a_1$ and $\rho=0$, $\a_4\le z\le\a_3$ of the symmetry axis will define the horizons of the upper and lower black holes, respectively (see Fig.~1a). The limiting procedure consists in tending $\a_2$ to $\a_1$, and $\a_4$ to $\a_3$ in formulas (\ref{EF}) and (\ref{ef_a}). The limit $\a_2=\a_1$, $\a_4=\a_3$ in (\ref{ef_a}) does not exhibit any difficulty; however, it is necessary to be careful while evaluating $f_1$ and $f_2$ in order not to omit any sign that these quantities might take. Then, in the extreme limit we readily get from (\ref{ef_a})
\be f_1=\epsilon_1\frac{(\beta_1-\a_1)(\beta_1-\a_3)}{\beta_2-\beta_1}, \quad f_2=\epsilon_2\frac{(\beta_2-\a_1)(\beta_2-\a_3)}{\beta_1-\beta_2}, \quad \epsilon_1=\pm1, \quad \epsilon_2=\pm1, \label{fi} \ee
the corresponding expressions for $e_1$ and $e_2$ being
\be e_1=-2\epsilon_1f_1\left[1+ \frac{(\epsilon_1-\epsilon_2)f_2}{\beta_1-\beta_2}\right], \quad e_2=-2\epsilon_2f_2\left[1+ \frac{(\epsilon_1-\epsilon_2)f_1}{\beta_1-\beta_2}\right]. \label{ei} \ee
As will be seen later, the concrete choice of $\epsilon_1$ and $\epsilon_2$ in (\ref{fi})--(\ref{ei}) determines whether the charges of black holes have equal or opposite signs.

The limiting procedure in the expressions defining Ernst potentials (\ref{EF}) requires the application of l'H\^opital's rule because, with $\a_2=\a_1$ and $\a_4=\a_3$, the determinants $E_\pm$ and $F$ vanish. However, we can write down immediately the final result for $\E$ and $\Phi$ if we observe that using of l'H\^opital's rule in (\ref{EF}) is equivalent to the application of Sibgatullin's method to the axis data (\ref{data}) under the supposition that the resulting solution will involve a pair of $\a$s of multiplicity two, something that was already done in the paper \cite{MSM} in a more general context of stationary electrovac spacetimes. Therefore, using the results of \cite{MSM} and taking into account the reality of the Ernst potentials in the electrostatic case, below we can write down the final form of $\E$ and $\Phi$ in the required extremal limit:
\bea &&{\cal E}=E_+/E_-, \quad \Phi=F/E_-, \nonumber\\ \nonumber\\
&&E_\pm=\left|\begin{array}{ccccc}
1 & \hspace{0.2cm}1 & \hspace{0.4cm}1 & \hspace{0.3cm}\dfrac{z-\a_1}{r_1} & \hspace{0.2cm}\dfrac{z-\a_3}{r_3} \vspace{0.2cm} \\ \pm1 &  & & & \vspace{0.2cm}\\ \pm1 & & & & \\\vspace{-0.35cm}  & &  & (A) & \\
0 & & & & \vspace{0.2cm}\\ 0 & & & & \\
\end{array}\right|, \nonumber\\ &&F=\left|\begin{array}{ccccc}
0 & \hspace{0.2cm}f(\a_1) & \hspace{0.1cm}f(\a_3) & \hspace{0.2cm}r_1^2\dfrac{\partial} {\partial\alpha_1}\left[\dfrac{f(\a_1)}{r_1}\right] & r_3^2\dfrac{\partial} {\partial\alpha_3}\left[\dfrac{f(\a_3)}{r_3}\right] \vspace{0.2cm} \\ -1 &  & & & \vspace{0.2cm} \\ -1 & & & & \\\vspace{-0.35cm}  & &  & (A) & \\
0 & & & & \vspace{0.2cm}\\ 0 & & & & \\
\end{array}\right|, \nonumber\\ \nonumber\\ &&A=\left(\begin{array}{cccc} \dfrac{r_1}{\alpha_1 -
\beta_1} & \dfrac{r_3}{\alpha_3 - \beta_1} & - \dfrac{r_1^2}{(\alpha_1 - \beta_1)^2}
& - \dfrac{r_3^2}{(\alpha_3 - \beta_1)^2}\vspace{0.25cm}\\
\vspace{0.25cm}\dfrac{r_1}{\alpha_1 - \beta_2} &
\dfrac{r_3}{\alpha_3 - \beta_2} & - \dfrac{r_1^2}{(\alpha_1 - \beta_2)^2}
& - \dfrac{r_3^2}{(\alpha_3 - \beta_2)^2}\\
\vspace{0.25cm}\dfrac{h_1(\a_1)}{\alpha_1 - \beta_1} &
\dfrac{h_1(\a_3)}{\alpha_3 - \beta_1} & r_1^2\dfrac{\partial} {\partial
\alpha_1}\left[\dfrac{h_1(\a_1)}
{(\alpha_1-\beta_1)r_1}\right] & r_3^2\dfrac{\partial} {\partial
\alpha_3}\left[\dfrac{h_1(\a_3)}
{(\alpha_3-\beta_1)r_3}\right]\\
\dfrac{h_2(\a_1)}{\alpha_1 - \beta_2} & \dfrac{h_2(\a_3)}{\alpha_3 -
\beta_2} & r_1^2\dfrac{\partial} {\partial
\alpha_1}\left[\dfrac{h_2(\a_1)}
{(\alpha_1-\beta_2)r_1}\right] &
r_3^2\dfrac{\partial} {\partial
\alpha_3}\left[\dfrac{h_2(\a_3)}
{(\alpha_3-\beta_2)r_3}\right]\\
\end{array}\right), \label{EF1}  \eea
where, according to the conventions of paper \cite{MSM}, partial differentiation in the determinant $F$ and in the matrix $A$ is applied only to the quantities $f(\a_n)$, $(\a_n-\beta_l)$ and functions $r_n$, the constants $e_l$, $\beta_l$, $f_l$ being assumed independent of $\a_1$ and $\a_3$; thus, for instance,
\be \dfrac{\partial} {\partial
\alpha_1}\left[\dfrac{h_1(\a_1)}
{(\alpha_1-\beta_1)r_1}\right]=h_1(\a_1)\left[\frac{z-\a_1}{(\a_1-\beta_1)r_1^3} -\frac{1}{(\a_1-\beta_1)^2r_1}\right]-\frac{2f_1}{(\a_1-\beta_1)r_1} \sum\limits_{k=1}^2\frac{f_k}{(\a_1-\beta_k)^2}. \label{deriv} \ee

The form of the corresponding metric coefficients $f$ and $\gamma$ entering Weyl's line element
\be
\d s^2=f^{-1}[e^{2\gamma}(\d\rho^2+\d z^2)+\rho^2\d\varphi^2]-f\d
t^2, \label{Weyl} \ee
is obtainable from formula (14) of \cite{MSM}, yielding
$$ f=\frac{E_+E_-+F^2}{E_-^2}, \quad e^{2\gamma}=\frac{E_+E_-+F^2}{K_0^2r_1^4r_3^4}, $$
\be K_0=\left|\begin{array}{cccc} \dfrac{1}{\alpha_1 -
\beta_1} & \dfrac{1}{\alpha_3 - \beta_1} & - \dfrac{1}{(\alpha_1 - \beta_1)^2}
& - \dfrac{1}{(\alpha_3 - \beta_1)^2}\vspace{0.25cm}\\
\vspace{0.25cm}\dfrac{1}{\alpha_1 - \beta_2} &
\dfrac{1}{\alpha_3 - \beta_2} & - \dfrac{1}{(\alpha_1 - \beta_2)^2}
& - \dfrac{1}{(\alpha_3 - \beta_2)^2}\\
\vspace{0.25cm}\dfrac{h_1(\a_1)}{\alpha_1 - \beta_1} &
\dfrac{h_1(\a_3)}{\alpha_3 - \beta_1} & \dfrac{\partial} {\partial
\alpha_1}\left[\dfrac{h_1(\a_1)}
{\alpha_1-\beta_1}\right] & \dfrac{\partial} {\partial
\alpha_3}\left[\dfrac{h_1(\a_3)}
{\alpha_3-\beta_1}\right]\\
\dfrac{h_2(\a_1)}{\alpha_1 - \beta_2} & \dfrac{h_2(\a_3)}{\alpha_3 -
\beta_2} & \dfrac{\partial} {\partial
\alpha_1}\left[\dfrac{h_2(\a_1)}
{\alpha_1-\beta_2}\right] &
\dfrac{\partial} {\partial
\alpha_3}\left[\dfrac{h_2(\a_3)}
{\alpha_3-\beta_2}\right]\\
\end{array}\right|. \label{fg_gen}  \ee

Therefore, we have found a representation of the extreme limit of the DRN solution in terms of $5\times5$ determinants. It should be pointed out that once the general expressions for the Ernst potentials and for the whole metric are obtained, these can be further reparametrized by employing any of the equivalent parameter sets that one might have at his disposal. At present moment we go on using the set $\{\a_n,\beta_l\}$ which, after the limiting procedure has been carried out, reduces to only four arbitrary constants, $\a_1$, $\a_3$, $\beta_1$, $\beta_2$, representing three physical quantities (recall that the sum of $\a_1$ and $\a_3$ can be assigned any real constant value). The form of $f_l$ and $e_l$ which enter the constant objects $h_l(\a_n)$ is defined in terms of $\a_n$ and $\beta_l$ by formulas (\ref{fi}) and (\ref{ei}). Our next major objective will be rewriting the obtained general expressions for the potentials $\E$, $\Phi$ and metric functions $f$, $\gamma$ in a simpler form by substituting formulas (\ref{fi}), (\ref{ei}) into (\ref{EF1}), (\ref{fg_gen}) and expanding the determinants.

\section{The general extreme solution in spheroidal coordinates}

Although the procedure of the expansion of determinants in formulas (\ref{EF1}) and (\ref{fg_gen}) involving the general values of $f_l$ and $e_l$ might look complicated at first glance, this is not really the case because the corresponding analytical computer processing does not exhibit much difficulty and leads straightforwardly to the desired result. We only mention that it is advantageous to choose the origin of coordinates in such a way that $\a_1+\a_3=0$ (see Fig.~1b), and also introduce prolate spheroidal coordinates $(x,y)$ via the formulas
\be \a_1=\a, \quad \a_3=-\a, \quad x=\frac{1}{2\a}(r_3+r_1), \quad y=\frac{1}{2\a}(r_3-r_1), \label{xy} \ee
where $\a$ is a positive real constant. The form of $r_1$, $r_3$ and $z$ in terms of $x$ and $y$ is
\be
r_1=\a(x-y), \quad r_3=\a(x+y), \quad z=\a xy. \label{zxy} \ee

In the new coordinates the line element (\ref{Weyl}) rewrites as
\be
\d s^2=\a^2f^{-1}\left[e^{2\gamma}(x^2-y^2)\left(\frac{\d x^2}{x^2-1}+\frac{\d y^2}{1-y^2}\right)+(x^2-1)(1-y^2)\d\varphi^2\right]-f\d
t^2. \label{Weyl_xy} \ee

Below we give the final expressions for $\E$, $\Phi$, $f$, $\gamma$ at which we have eventually arrived after the substitution of (\ref{fi}), (\ref{ei}) into (\ref{EF1}), (\ref{fg_gen}) and simplifications:
\bea \E&=&\frac{E_+}{E_-}, \quad \Phi=\frac{\epsilon_1(\beta_1-\beta_2)F}{E_-}, \quad f=\frac{N}{E_-^2}, \quad e^{2\gamma}=\frac{N}{4N_2^4}, \nonumber\\ E_\pm&=&(\epsilon_1-\epsilon_2) [\a(\beta_1-\beta_2)x-(\a^2-\beta_1\beta_2)y\pm(\beta_1^2-\a^2)] \nonumber\\ &\times&[\a(\beta_1-\beta_2)x+(\a^2-\beta_1\beta_2)y\pm(\a^2-\beta_2^2)]N_1 \nonumber\\ &+&(\epsilon_1+\epsilon_2)(\beta_1-\beta_2)^2 [\a^2(x^2-y^2)\pm\a(\beta_1+\beta_2)x\mp(\a^2+\beta_1\beta_2)y]N_2, \nonumber\\ F&=&(\epsilon_1-\epsilon_2)[\a(2\a^2-\beta_1^2-\beta_2^2)x-(\beta_1+\beta_2) (\a^2-\beta_1\beta_2)y]N_1 \nonumber\\ &+&(\epsilon_1+\epsilon_2) (\beta_1-\beta_2)[-\a(\beta_1+\beta_2)x+(\a^2+\beta_1\beta_2)y]N_2, \nonumber\\ N&=&(\epsilon_1-\epsilon_2)^2N_1^4+(\epsilon_1+\epsilon_2)^2N_2^4, \nonumber\\ N_1&=&\a^2(\beta_1-\beta_2)^2x^2-(\a^2-\beta_1\beta_2)^2y^2 +(\a^2-\beta_1^2)(\a^2-\beta_2^2), \nonumber\\ N_2&=&\a^2(\beta_1-\beta_2)^2(x^2-y^2). \label{metric_MPB} \eea

Formulas (\ref{metric_MPB}) are fully equivalent to formulas (\ref{fi}), (\ref{ei}), (\ref{EF1}), (\ref{fg_gen}) of the previous section, but due to their concise form they are of course by far more advantageous as a representation of the general extreme DRN solution than the determinantal expressions. In our opinion, the most proper name for the field defined by (\ref{metric_MPB}) would be {\it the combined Majumdar--Papapetrou--Bonnor solution}, and in what follows we will justify this name by considering two particular specializations of the solution (\ref{metric_MPB}) arising from two different relations between the quantities $\epsilon_1$ and $\epsilon_2$.

\subsection{The Majumdar--Papapetrou two--body solution}

When $\epsilon_1=\epsilon_2=\epsilon$, $\epsilon=\pm1$, we obtain from (\ref{metric_MPB})
\bea \E&=&\frac{E_+}{E_-}=\frac{\a^2(x^2-y^2)+\a(\beta_1+\beta_2)x-(\a^2+\beta_1\beta_2)y} {\a^2(x^2-y^2)-\a(\beta_1+\beta_2)x+(\a^2+\beta_1\beta_2)y}, \nonumber\\
\Phi&=&\frac{F}{E_-}=\frac{\epsilon[-\a(\beta_1+\beta_2)x+(\a^2+\beta_1\beta_2)y]} {\a^2(x^2-y^2)-\a(\beta_1+\beta_2)x+(\a^2+\beta_1\beta_2)y}, \label{EF_MP} \eea
and these Ernst potentials define the Majumdar--Papapetrou solution for two extreme Reissner--Nordstr\"om black holes. To see this better, let us also write out the corresponding metric functions $f$ and $\gamma$:
\be f=\frac{\a^4(x^2-y^2)^2}{E_-^2}, \quad e^{2\gamma}=1, \label{f_MP}  \ee
whence it follows that the solution (\ref{EF_MP}) is conformastatic \cite{Boo}. Moreover, in the particular case under consideration one finds from (\ref{MQ_t}), (\ref{fi}) and (\ref{ei}) that
\be e_1+e_2=-2\epsilon(f_1+f_2) \quad \Longleftrightarrow \quad M_1+M_2=\epsilon(Q_1+Q_2), \label{MQ} \ee
$M_i$ being the Komar masses and $Q_i$ the Komar charges of black holes. To show that this implies the equalities $Q_i=\epsilon M_i$ characterizing the Majumdar--Papapetrou solution, one has to make use of the quantities $\sigma_1$ and $\sigma_2$ of Ref.~\cite{Man}. Then, expressing the separation $R$ from the equation $\sigma_1=0$ and substituting the result into the equation $\sigma_2=0$, one can easily solve the latter for $M_2$; this $M_2$ being substituted into (\ref{MQ}) finally gives $Q_1=\epsilon M_1$. In analogy one obtains $Q_2=\epsilon M_2$. Therefore, the equivalence of the solution (\ref{EF_MP}) to the $N=2$ specialization of the Majumdar--Papapetrou spacetime is evident.

The results of the paper \cite{Man} also permit one to relate the parameters $\beta_1$ and $\beta_2$ entering (\ref{EF_MP}) to the Komar masses and to the distance $\a=R/2$ of each extreme source from the origin of coordinates:
\be \beta_1+\beta_2=-M_1-M_2, \quad \beta_1\beta_2=-\a(\a+M_1-M_2) \label{b1b2_MP} \ee
(the subscripts 1 and 2 label, respectively, the lower and upper constituents), so that the polynomials $E_\pm$ and $F$ defining a representation of the two--body Majumdar--Papapetrou solution in prolate spheroidal coordinates can be rewritten as
\bea E_\pm&=&\a(x^2-y^2)\mp (M_1+M_2)x\pm (M_1-M_2)y, \nonumber\\
F&=&\epsilon[(M_1+M_2)x-(M_1-M_2)y], \label{Epm} \eea
thus allowing for a slight simplification of formulae (\ref{EF_MP}) and (\ref{f_MP}).

It is worth noting that, as it follows from a comprehensive analysis of MP spacetimes carried out by Hartle and Hawking \cite{HHa}, Eqs.~(\ref{EF_MP}), (\ref{f_MP}) describe a regular geometry of two charged extreme black holes only in the case $M_i>0$. If one of the masses is a negative quantity, the solution develops naked singularities off the symmetry axis, even when the total mass of the system is positive. The above said is well illustrated by Fig.~2 in which we have plotted singular surfaces for the following two particular parameter choices: (i) $M_1=2$, $M_2=-1$, $\a=3$, and (ii) $M_1=-2$, $M_2=-1$, $\a=3$. For that reason, throughout the present paper only the constituents with positive Komar masses are referred to as black holes.

\subsection{Bonnor's solution: the electrostatic analog of Kerr--NUT spacetime}

The other relation between $\epsilon_1$ and $\epsilon_2$ is $\epsilon_1=-\epsilon_2=\epsilon$, $\epsilon=\pm1$, and in this case we get from (\ref{metric_MPB})
\bea \E&=&\frac{E_+}{E_-}, \quad \Phi=\frac{\epsilon(\beta_1-\beta_2)F}{E_-}, \quad f=\frac{N^2}{E_-^2}, \quad e^{2\gamma}=\frac{N^4}{\a^8(\beta_1-\beta_2)^8(x^2-y^2)^4}, \nonumber\\ E_\pm&=&[\a(\beta_1-\beta_2)x-(\a^2-\beta_1\beta_2)y\pm(\beta_1^2-\a^2)] \nonumber\\ &\times&[\a(\beta_1-\beta_2)x+(\a^2-\beta_1\beta_2)y\pm(\a^2-\beta_2^2)], \nonumber\\ F&=&\a(2\a^2-\beta_1^2-\beta_2^2)x-(\beta_1+\beta_2) (\a^2-\beta_1\beta_2)y, \nonumber\\ N&=&\a^2(\beta_1-\beta_2)^2x^2-(\a^2-\beta_1\beta_2)^2y^2 +(\a^2-\beta_1^2)(\a^2-\beta_2^2). \label{EF_B} \eea

This is the three--parameter electrostatic solution first obtained by Bonnor \cite{Bon2}, though in a different form and not revealing that it was the electrostatic analog of the well--known Kerr--NUT spacetime \cite{DNe}, i.e., obtainable from the latter via Bonnor's procedure of a complex continuation of the parameters \cite{Bon3}. In the particular case  $2\a^2=\beta_1^2+\beta_2^2$ it further reduces to the solution for a massive electric/magnetic dipole \cite{Bon4}. The relation between our parameters $\a$, $\beta_1$, $\beta_2$ and the constants $a$, $A$, $B$, $C$ of the original Bonnor's paper~\cite{Bon2} is given by the formulae
\be
\a=a, \quad \beta_1=a(A+B), \quad \beta_2=-a(A+C), \quad BC=A^2-1. \label{rel_ourB} \ee

The simplest way to show how (\ref{EF_B}) can be constructed from the Kerr--NUT solution is to use the expression of the corresponding Ernst potential from Ref.~\cite{MMR} (formulae (3) of \cite{MMR} with $q=b=0$) and the Bonnor representation of the axisymmetric electrostatic problem (equations (2.9) and (2.10) of \cite{Bon2}). Then, after changing $a\to ia$, $\nu\to i\nu$ in the Ernst potential $\E$ of the Kerr--NUT solution, and also in its complex conjugate expression $\bar\E$, one will arrive at two real potentials $X$ and $Y$ whose product will give precisely the electrostatic Ernst potential $\E$ from (\ref{EF_B}), while the difference of $X$ and $Y$ will give the doubled value of the potential $\Phi$ from (\ref{EF_B}). The relation of the parameters $\a$, $\beta_1$, $\beta_2$ entering (\ref{EF_B}) to the Kerr--NUT parameters $m$, $a$, $\nu$ (the latter two after their complex continuation) is the following:
\be \a=\sqrt{m^2+a^2-\nu^2}, \quad \beta_1=-m+a+\nu, \quad \beta_2=-m-a-\nu. \label{DNpar} \ee

It should be pointed out that solution (\ref{EF_B}) has important aspects of its physical interpretation that need to be clarified. In the original paper \cite{Bon2} it was only referred to as describing a mass endowed with both electric charge and dipole moment. At the same time, during the last decade there appeared some new interesting results shedding additional light on the interpretation of Bonnor's solutions. Thus, for instance, in Ref.~\cite{Emp} the electric/magnetic dipole spacetime \cite{Bon4} was shown by Emparan to arise from two charged extreme black holes with equal masses and equal but opposite charges, and later on this result was generalized by Liang and Teo \cite{LTe} to the case of the non--zero net charge of black--hole constituents. However, whereas in the paper \cite{Emp} a correct relation between the charges and masses of extreme black holes was established, the authors of~\cite{LTe} presented erroneous relations between the masses and charges of the constituents\footnote{For example, a correct expression for the masses of extreme black holes in the absence of dilatonic field is $m\pm(l/a)\sqrt{m^2+a^2-l^2}$ (in notations of \cite{LTe}), and not $m\pm l$ as was given by Liang and Teo.} and, besides, they did not identify the electrostatic analog of the Kerr--NUT metric as the Bonnor three--parameter solution \cite{Bon2} (but, importantly, they correctly indicated, on the one hand, the inequality of masses, and, on the other hand, the inequality of charges in their two--body configurations). A crucial mistake made by Liang and Teo was to assume that the extreme black holes in their Einstein--Maxwell dihole solution were of the MP type, i.e., verifying $Q_i^2=M_i^2$, with the only distinction that the charges had to have opposite signs.

The correct formulae relating the charges and masses of the extreme black--hole constituents of the solution (\ref{EF_B}) are readily obtainable by equating the quantities $\sigma_1$ and $\sigma_2$ of the paper~\cite{Man} to zero and excluding the MP case; we thus get
\be Q_1=\epsilon M_1\sqrt{\frac{(R+M_2)^2-M_1^2}{(R-M_2)^2-M_1^2}}, \quad  Q_2=-\epsilon M_2\sqrt{\frac{(R+M_1)^2-M_2^2}{(R-M_1)^2-M_2^2}}, \label{Qi_B} \ee
whence the Liang--Teo conjectured relations follow only in the limit $R\to\infty$ of infinite separation. One can also see that Emparan's result for the Bonnor dipole solution is recovered from (\ref{Qi_B}) when $M_1=M_2$.

Formulae (\ref{Qi_B}) clearly demonstrate that the charges of extreme constituents in the solution (\ref{EF_B}) always have opposite signs, provided that $M_i>0$. Another important conclusion following from (\ref{Qi_B}) is that the absolute values of these charges are greater than the respective values of the masses, $|Q_i|>M_i$. This curious property of the non--isolated  extreme charged black holes was first established by Emparan in the case of identical (up to the sign of the charge) black--hole constituents \cite{Emp}. The present paper extends Emparan's discovery to the case of extreme constituents of the non--MP type with unequal positive masses. Mention here that the negative masses of the constituents are likely to be excluded on the same grounds as in the MP case -- they are accompanied by naked singularities, and in Fig.~3 we have plotted singular surfaces for the same choices of the parameters $M_i$, $\a$ which were earlier used for plotting Fig.~2.

We would like to conclude this section by noting that the physical quantities are related to the parameters $\a$, $\beta_1$, $\beta_2$ of Bonnor's solution by the formulae
\bea M_1&=&\frac{(\a+\beta_1)(\a+\beta_2)(\beta_1+\beta_2-2\a)} {2(\a^2-\beta_1\beta_2)}, \nonumber\\ M_2&=&\frac{(\a-\beta_1)(\a-\beta_2)(\beta_1+\beta_2+2\a)} {2(\a^2-\beta_1\beta_2)}, \nonumber\\ Q_1^2&=&M_1^2\frac{(\beta_1+\beta_2-2\a)^2} {(\beta_1-\beta_2)^2}, \quad Q_2^2=M_2^2\frac{(\beta_1-\beta_2)^2}{(\beta_1+\beta_2+2\a)^2}, \eea
while the solution (\ref{EF_B}) can be rewritten in terms of the physical parameters $R$, $M_1$, $M_2$ by means of the substitutions
\bea \a&=&\frac{1}{2}R, \quad \beta_1=-\frac{1}{2}(D+M_1+M_2), \quad \beta_2=\frac{1}{2}(D-M_1-M_2), \nonumber\\  D&=&\sqrt{\frac{(R-M_1-M_2)[(R+M_1)^2-M_2^2]}{R-M_1+M_2}}. \label{sub} \eea
The above relations between the parameter sets are a direct consequence of the results of Ref.~\cite{Man}.

\section{Discussion and conclusions}

In the present paper we have elaborated the general extreme limit of two non--rotating charged black holes starting from the five--parameter DRN solution in the canonical representation. This limit is represented by the combined Majumdar--Papapetrou--Bonnor field whose particular specializations are the two--body MP spacetime written in prolate spheroidal coordinates (when the charges of black holes have equal signs) and Bonnor's electrostatic solution (the case of oppositely charged masses). The extreme limit involves in general three arbitrary real parameters which can be associated with the Komar masses of the black holes and the separation distance. At the same time, whereas the extreme constituents of the MP solution are in equilibrium which is distance--independent, the equilibrium states within the Bonnor solution do not exist, as was shown by Liang and Teo \cite{LTe}, so the constituents in this particular solution are kept apart from each other by a supporting strut \cite{Isr}. An important characteristic feature of the extreme black holes in the Bonnor solution demonstrated by us is the inequality $|Q_i|>M_i$ between the individual charges and masses, and it generalizes Emparan's result previously obtained for a particular case of identical constituents \cite{Emp}. This means, taking into account the equality $Q_i^2=M_i^2$ valid for extreme constituents of the MP solution, that in the two--body systems of charged extreme black holes the absolute value of the individual Komar charge can never be less than the respective individual Komar mass.

It is worth noting that formulas (\ref{EF_MP})--(\ref{Epm}) describing the two--black--hole MP spacetime in prolate spheroidal coordinates are well suited for the illustration that the extreme black--hole constituents are not the only axisymmetric objects in equilibrium covered by the MP class of solutions. Indeed, while working with the $x$ and $y$ coordinates, it is quite customary to explore both the prolate and oblate variants of the coordinate system, so in (\ref{EF_MP})--(\ref{Epm}) we can pass to the latter (oblate) variant by the transformation
\be x\to ix, \quad \a\to -i\a, \quad y\to y, \label{transf} \ee
which, however, should not affect the reality of the Ernst potentials $\E$ and $\Phi$. From (\ref{Epm}) it is easy to see that the condition for $\E$ and $\Phi$ to continue to stay real under (\ref{transf}) is the equality of the masses $M_1$ and $M_2$. Therefore, after setting $M_1=M_2={\textstyle\frac12}M$ and passing in the MP solution for two extreme black holes to the oblate spheroidal coordinates by means of (\ref{transf}) we arrive at the following transformed metric and electric $\Phi$ potential:
\bea
\d s^2&=&\a^2f^{-1}\left[(x^2+y^2)\left(\frac{\d x^2}{x^2+1}+\frac{\d y^2}{1-y^2}\right)+(x^2+1)(1-y^2)\d\varphi^2\right]-f\d
t^2, \nonumber\\ f&=&\frac{\a^2(x^2+y^2)^2}{[\a(x^2+y^2)+Mx]^2}, \quad  \Phi=\frac{\epsilon Mx}{\a(x^2+y^2)+Mx}. \label{ring} \eea
The above formulas represent what we would call the simplest MP ring, whose mass and charge are both equal to $M$ in absolute value, and in the paper \cite{PRW} this solution arose as a particular static limit of a more general stationary electrovac spacetime. For us, the most interesting point about the solution (\ref{ring}) is that although its mass and charge satisfy, like in the case of an isolated extreme RN black hole, the relation $Q^2=M^2$, it is not already spherically symmetric because it possesses higher multipole moments \cite{PRW}.

If we now return to Weyl's cylindrical coordinates and superpose the ring solution with the extreme RN field, we shall arrive at a particular $N=3$ specialization of the MP spacetime of the form
\bea
\d s^2&=&(1+\psi_r+\psi_h)^2(\d\rho^2+\d z^2+\rho^2\d\varphi^2) -(1+\psi_r+\psi_h)^{-2}\d
t^2, \quad \Phi=\frac{\epsilon(\psi_r+\psi_h)}{1+\psi_r+\psi_h}, \nonumber\\ \psi_r&=&\frac{M}{2}\left(\frac{1}{\sqrt{\rho^2+(z+ia)^2}} +\frac{1}{\sqrt{\rho^2+(z-ia)^2}}\right), \quad \psi_h=\frac{m}{\sqrt{\rho^2+(z-b)^2}}, \label{super} \eea
where $m$ is the mass of the black hole and $b$ its position on the symmetry axis (the subindices $r$ and $h$ refer to the ring and the black hole, respectively). It is remarkable that the ring and and the extreme black hole are forming a {\it distance--independent} equilibrium configuration which is similar to the one of the two--black--hole MP solution. Moreover, a trivial modification of the potential $\psi_h$ in (\ref{super}) leads to the equilibrium states between two MP rings that are also distance--independent! Therefore, actually we have at hand three different geometries representing the two--body equilibrium configurations independent of separation and involving the equality $Q_i=\epsilon M_i$. In our opinion, this might be regarded as a possible indication that the classical equilibrium condition $M_1M_2=Q_1Q_2$ for charged point particles must also hold in general relativity if the effects of the induced electric field \cite{HRu,MRu} are not taken into account.

\section*{Acknowledgments}

We would like to thank Prof. W. G. Unruh for valuable remarks on an earlier version of the paper. We also are grateful to the referees for various useful comments and suggestions. Our research was supported by Project FIS2006-05319 from Ministerio de Ciencia y Tecnolog\'\i a, Spain, and by the Junta de Castilla y Le\'on under the ``Programa de Financiaci\'on de la Actividad Investigadora del Grupo de Excelencia GR-234'', Spain.

\vspace{5cm}

\begin{figure}[htb]
\centerline{\epsfysize=100mm\epsffile{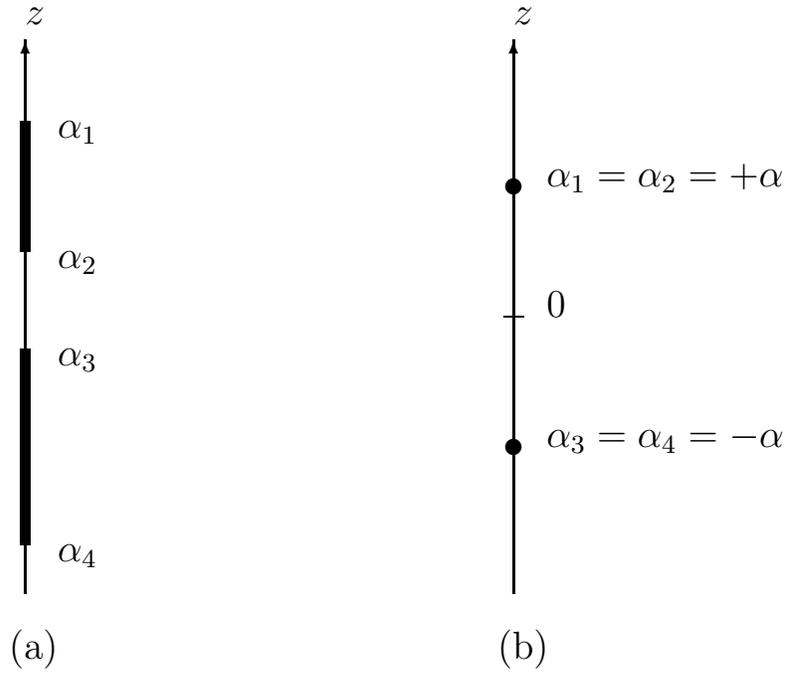}} \caption{Location of the black--hole constituents on the symmetry axis: (a) the subextreme case; (b) the case of two extreme constituents.}
\end{figure}

\begin{figure}[htb]
\centerline{\epsfysize=85mm\epsffile{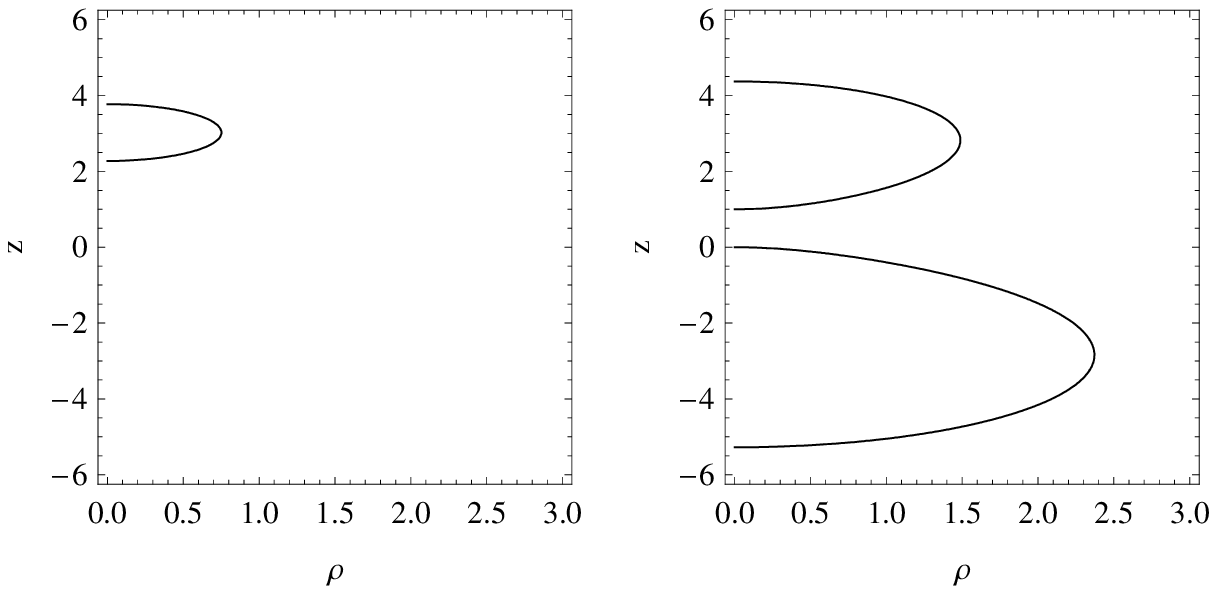}}
\caption{Formation of the singular spheroidal surfaces in the case of the MP two--body solution when one or two constituents have negative masses.}
\label{fig:2}       
\end{figure}

\begin{figure}[htb]
\centerline{\epsfysize=85mm\epsffile{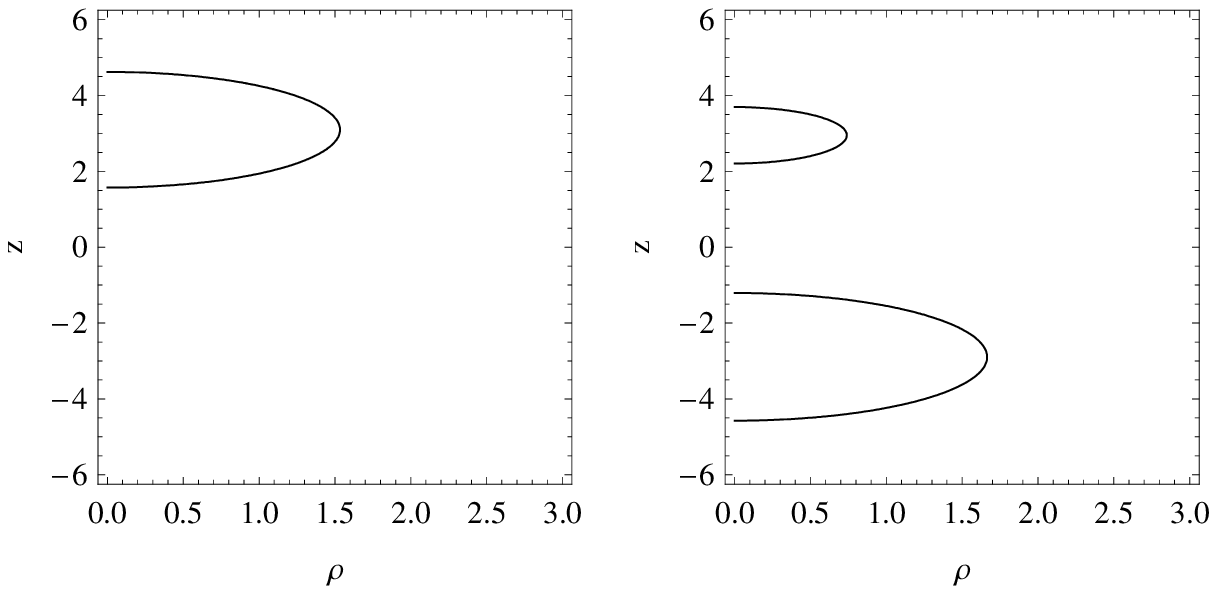}}
\caption{Formation of the singular spheroidal surfaces in the case of the Bonnor solution when one or two constituents have negative masses.}
\label{fig:3}       
\end{figure}

\end{document}